\documentclass[journal=jacsat,manuscript=article]{achemso}

\usepackage[version=3]{mhchem} 

\usepackage{physics}
\usepackage{xargs}
\usepackage{booktabs}
\usepackage[dvipsnames]{xcolor}
\usepackage{chemformula} 
\usepackage{rotating} 

\newcommand{\cfour}{\textsc{cfour }} 

\newcommand{\oper}[1]{\hat{#1}} 

\newcommand{\hD}{\oper{h}_{\rm D}}
\newcommand{\gop}{\oper{g}}
\newcommand{\HDC}{\oper{H}_{\rm DC}}

\newcommand{\mat}[1]{\mathbf{#1}}

\newcommand{\mW}{\mat{W}}
\newcommand{\mWLL}{\mW^{\rm LL}}
\newcommand{\mWLS}{\mW^{\rm LS}}
\newcommand{\mWSL}{\mW^{\rm SL}}
\newcommand{\mWSS}{\mW^{\rm SS}}

\newcommand{\mL}{\mat{L}}
\newcommand{\mLLP}[1]{\mL^{{\rm L},#1}}
\newcommand{\mLSP}[1]{\mL^{{\rm S},#1}}

\newcommand{\DLL}[1]{D_{#1}^{\rm LL}}
\newcommand{\DLS}[1]{D_{#1}^{\rm LS}}
\newcommand{\DSL}[1]{D_{#1}^{\rm SL}}
\newcommand{\DSS}[1]{D_{#1}^{\rm SS}}

\newcommand{\WLL}[1]{W_{#1}^{\rm LL}}
\newcommand{\WLS}[1]{W_{#1}^{\rm LS}}
\newcommand{\WSL}[1]{W_{#1}^{\rm SL}}
\newcommand{\WSS}[1]{W_{#1}^{\rm SS}}

\newcommand{\CL}[1]{C_{#1}^{\rm L}}
\newcommand{\CS}[1]{C_{#1}^{\rm S}}

\newcommand{\V}{V}
\newcommand{\T}{T}
\newcommand{\vp}{\Vec{p}}
\newcommand{\vs}{\Vec{\sigma}}
\newcommand{\sx}{{\sigma_x}}
\newcommand{\sy}{{\sigma_y}}
\newcommand{\sz}{{\sigma_z}}
\newcommand{\vA}{\oper{\Vec{A}}}
\newcommand{\vB}{\oper{\Vec{B}}}

\newcommand{\psiL}{\psi^{\rm L}}
\newcommand{\psiS}{\psi^{\rm S}}
\newcommand{\phiL}{\phi^{\rm L}}

\newcommand{\braS}[1]{\bra{\psiS_{#1}}}
\newcommand{\ketS}[1]{\ket{\psiS_{#1}}}
\newcommand{\braL}[1]{\bra{\psiL_{#1}}}
\newcommand{\ketL}[1]{\ket{\psiL_{#1}}}
\newcommand{\brapL}[1]{\bra{\phiL_{#1}}}
\newcommand{\ketpL}[1]{\ket{\phiL_{#1}}}
\newcommand{\brapsi}[1]{\bra{\psi_{#1}}}
\newcommand{\ketpsi}[1]{\ket{\psi_{#1}}}
\newcommand{\ketLS}[1]{\ket{\psi^{\rm (L,S)}_{#1}}}

\newcommand{\twoelint}[4]{\big({#1}{#2}\big|{#3}{#4}\big)}
\newcommand{\twoelintpsi}[4]{\twoelint{\psi_{#1}}{\psi_{#2}}{\psi_{#3}}{\psi_{#4}}}

\newcommand{\LLLL}[4]{\twoelint{\psiL_{#1}}{\psiL_{#2}}{\psiL_{#3}}{\psiL_{#4}}}
\newcommand{\LLSS}[4]{\twoelint{\psiL_{#1}}{\psiL_{#2}}{\psiS_{#3}}{\psiS_{#4}}}
\newcommand{\SSLL}[4]{\twoelint{\psiS_{#1}}{\psiS_{#2}}{\psiL_{#3}}{\psiL_{#4}}}
\newcommand{\SSSS}[4]{\twoelint{\psiS_{#1}}{\psiS_{#2}}{\psiS_{#3}}{\psiS_{#4}}}

\newcommand{\LLPP}[4]{\big(\psiL_{#1}\psiL_{#2}\big|\big\{\vs\cdot\vp \phiL_{#3}\big\} \big\{\vs\cdot\vp \phiL_{#4}\big\} \big)}
\newcommand{\PPLL}[4]{\big(\big\{\vs\cdot\vp \phiL_{#1}\big\} \big\{\vs\cdot\vp \phiL_{#2}\big\} \big|\psiL_{#3}\psiL_{#4}\big)}
\newcommand{\PPPP}[4]{\big(\big\{\vs\cdot\vp \phiL_{#1}\big\} \big\{\vs\cdot\vp \phiL_{#2}\big\} \big|\big\{\vs\cdot\vp\phiL_{#3}\big\} \big\{\vs\cdot\vp \phiL_{#4}\big\} \big)}

\newcommand{\intLLLL}{\twoelint{{\rm L}}{{\rm L}}{{\rm L}}{{\rm L}}}
\newcommand{\intLLSS}{\twoelint{{\rm L}}{{\rm L}}{{\rm S}}{{\rm S}}}
\newcommand{\intSSLL}{\twoelint{{\rm S}}{{\rm S}}{{\rm L}}{{\rm L}}}
\newcommand{\intSSSS}{\twoelint{{\rm S}}{{\rm S}}{{\rm S}}{{\rm S}}}

\newcommand{\LLSSSF}[4]{\big(\psiL_{#1}\psiL_{#2}\big|\big\{\vp\phiL_{#3}\big\}\cdot\big\{\vp\phiL_{#4}\big\}\big)}
\newcommand{\SSLLSF}[4]{\big(\big\{\vp\phiL_{#1}\big\}\cdot\big\{\vp\phiL_{#2}\big\}\big|\psiL_{#3}\psiL_{#4}\big)}
\newcommand{\SSSSSF}[4]{\big(\big\{\vp\phiL_{#1}\big\}\cdot\big\{\vp\phiL_{#2}\big\}\big|\big\{\vp\phiL_{#3}\big\}\cdot\big\{\vp\phiL_{#4}\big\}\big)}

\newcommand{\LLSSSFAO}[4]{\big({#1}{#2}\big|\big\{\vp{#3}\big\}\cdot\big\{\vp{#4}\big\}\big)}
\newcommand{\SSLLSFAO}[4]{\big(\big\{\vp{#1}\big\}\cdot\big\{\vp{#2}\big\}\big|{#3}{#4}\big)}
\newcommand{\SSSSSFAO}[4]{\big(\big\{\vp{#1}\big\}\cdot\big\{\vp{#2}\big\}\big|\big\{\vp{#3}\big\}\cdot\big\{\vp{#4}\big\}\big)}

\newcommand{\cdthr}{\tau}
\newcommand{\ncv}{N_{\rm CV}}

\author{Tereza Uhl\'i\v{r}ov\'a}
\affiliation{Department Chemie, Johannes Gutenberg-Universität Mainz, Duesbergweg 10-14, D-55128 Mainz, Germany}
\author{Davide Cianchino}
\author{Tommaso Nottoli}
\author{Filippo Lipparini}
\email{filippo.lipparini@unipi.it}
\affiliation{Dipartimento di Chimica e Chimica Industriale, Universit\`a di Pisa, Via G. Moruzzi 13, I-56124 Pisa, Italy}
\author{J\"urgen Gauss}
\email{gauss@uni-mainz.de}
\affiliation{Department Chemie, Johannes Gutenberg-Universität Mainz, Duesbergweg 10-14, D-55128 Mainz, Germany}

\title{
Cholesky decomposition in
spin-free Dirac-Coulomb
coupled-cluster
calculations
}

\begin{document}


\begin{tocentry}






\includegraphics[width=\textwidth]{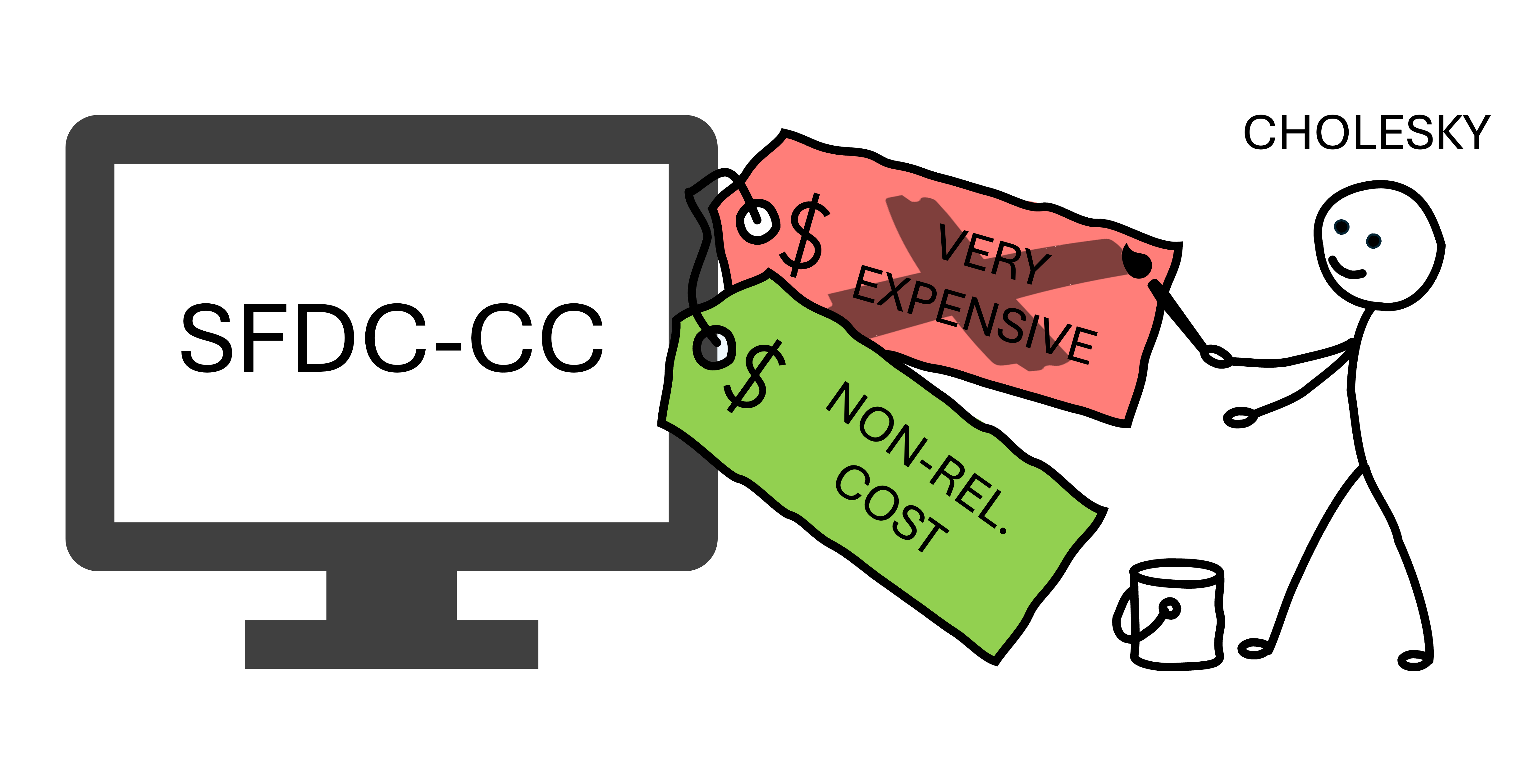}

\end{tocentry}


\begin{abstract}
We present an implementation for the use of Cholesky decomposition (CD) of two-electron integrals within the spin-free Dirac-Coulomb (SFDC) scheme that enables to perform
high-accuracy coupled-cluster (CC) calculations at costs almost comparable to those of their non-relativistic counterparts.
While for non-relativistic CC calculations atomic-orbital (AO) based algorithms, due to their significantly reduced disk-space requirements, are the key to efficient large-scale computations, such algorithms are less advantageous in the SFDC case due to their increased computational cost on that case. Here, molecular-orbital (MO) based algorithms exploiting the CD of the two-electron integrals allow to reduce disk-space requirements, 
and lead to computational cost in the CC step that are more or less the same as in the non-relativistic case. The only remaining overhead in a CD-SFDC-CC
calculation are due to the need to compute additional two-electron integrals, the somewhat higher cost of the Hartree-Fock calculation in the SFDC case,
and additional cost in the transformation of the Cholesky vectors from the AO to the MO representation.
However, these additional costs typically amount to less than
5--15 \% of the total wall time
and are thus acceptable. 
We illustrate the efficiency of our CD scheme for SFDC-CC calculations on a series of illustrative calculations
for the ${\rm X(CO)}_4$ molecules with ${\rm X} = {\rm Ni}, {\rm Pd}, {\rm Pt}$.
\end{abstract}


\section{Introduction}

Since its first wider recognition in the late 70s,\cite{Pitzer1979,Pyykko1978}
the field of relativistic quantum chemistry has been rapidly developing. Nowadays, the importance of including relativistic effects in high-accuracy calculations and in particular for the treatment of heavy-elements compounds has been well recognized
and a great variety of methods and programs are available for such relativistic quantum-chemical calculations (see the following reviews\cite{Pyykko1988,Saue2011,Autschbach2012,Fleig2012,Cheng2014b,Cremer2014} and references therein).

The usual starting point of relativistic quantum-chemical calculations is the no-pair Dirac-Coulomb (DC) Hamiltonian,\cite{Dyall2007,Sucher1980} which consists of the fully relativistic one-electron Dirac Hamiltonian
and of the non-relativistic Coulomb potential for the two-electron part. Relativistic corrections to the electron-electron repulsion can be further added by including Breit interactions in the Hamiltonian.\cite{Breit1929}
The most rigorous treatment would be then a direct DC calculation with a four-component scheme.\cite{Visscher1995,Visscher1996,Fleig2007,Nataraj2010,Sorensen2011}
However, due to spin-orbit (SO) coupling, this approach requires the use of complex algebra and can be only done with reduced exploitation of symmetry. It thus is computationally very demanding and a so-called four-component coupled-cluster (CC) singles and doubles (CCSD) calculation thus is typically about 30 times more expensive than its non-relativistic counterpart.\cite{Visscher1995}
In addition, due to the need to use complex algebra, one cannot simply exploit existing non-relativistic implementations of CC methods; new programs need to be developed.
Much effort has been therefore invested into the development of alternative and more cost-effective approaches.
Among those,
the Douglas-Kroll-Hess (DKH) approach,
\cite{Douglas1974,Hess1986,Jansen1989,Wolf2002,Reiher2004a,Reiher2004b,Peng2009}
zeroth-order regular approximation (ZORA),
\cite{Chang1986,Heully1986,Lenthe1993,Lenthe1994,Sadlej1995,Lenthe1996}
perturbative approaches such as the so-called direct perturbation theory (DPT),
\cite{Rutkowski1986a,Rutkowski1986b,Rutkowski1986c,Kutzelnigg1989,Kutzelnigg1990,Kutzelnigg1995,Kutzelnigg1996,Rutkowski1996,Stopkowicz2011}
approximate two-component approaches such as the exact two-component (X2C) and its many variants,
\cite{Dyall1997,Kutzelnigg2005,Liu2006,Filatov2007,Sikema2009,Liu2009,Zou2011,Cheng2011b,Cheng2014a,Cheng2018,Franzke2018,Verma2016}
and spin-free DC (SF, SFDC)
\cite{Visscher2000,Fleig2005,Knecht2008,Cheng2011a,Cheng2013}
schemes are the most popular ones.

The SFDC approximation is a well-established scheme
that is motivated by the distinct origin, manifestations, and different orders of magnitude of the scalar-relativistic and SO effects, and by the different computational and methodological requirements for their treatment.
As suggested by Dyall\cite{Dyall1994,Stanton1984}
and pursued by many since,\cite{Visscher2000,Fleig2005,Knecht2008,Cheng2011a,Cheng2013}
one can rigorously separate the SF and SO part and neglect the latter, achieving a rigorous treatment of scalar-relativistic effects while, at the same time, retaining a spin-symmetry conserving Hamiltonian. Consequently, one can use real algebra, the usual point-group as well as spin symmetries exactly as in the non-relativistic treatment.
In addition, the SFDC scheme allows one to use already existing non-relativistic quantum-chemistry codes with modification only in the integral evaluation and in the Hartree-Fock (HF) part, while the CC treatment (and more generally the electron-correlation treatment) is identical apart from a more involved transformation of the molecular integrals from the AO to the MO representation. However, the statement that electron-correlation treatments (at CC or other levels) are unchanged in costs when moving from the non-relativistic to the SFDC case holds only for electron-correlation treatments using MO based algorithms.
Unfortunately, such MO based algorithms require one to assemble integrals with four indices referring to virtual MOs, which requires $V^4$ words of disk space with $V$ as the number of virtual orbitals. Such an amount of disk space is generally either not available or renders a corresponding computation heavily I/O bound.

In the non-relativistic case, this problem can be circumvented by using AO based algorithms instead. Unfortunately, this strategy is less advantageous in SFDC calculations due to the increased number of two-electron integrals to be handled at the AO level. In practice, SFDC-CC calculations on medium to large system are thus much more expensive than their non-relativistic counterparts.

In this contribution, we use the Cholesky decomposition (CD) of the two-electron integrals\cite{Beebe1977,Koch2003,Aquilante2007,Aquilante2008a,Aquilante2008b,Epifanovsky2013,Piccardo2017,Nottoli2021a,Nottoli2021b,Banerjee2023,Zhang2024}
to overcome the limitations in large-scale SFDC-CC computations. The underlying idea\cite{Beebe1977,Koch2003} of CD is to use an auxiliary basis that is constructed on-the-fly to decompose the four-index matrix of the two-electron repulsion integrals (ERI matrix) into a three-index array (consisting of the so-called Cholesky vectors (CVs)), and thereby to reduce the computational requirements.
In addition, one usually introduces an \textit{a priori} threshold for the calculation of the CVs that allows for a rigorous error control. Consequently, only a small fraction of all possible ERIs needs to be calculated and hence a significant amount of computational time and space is saved.\cite{Koch2003}
Moreover, CD of ERIs can be relatively easily parallelized to further speed up the calculation.
CD of ERIs has been already implemented into numerous programs at various levels of non-relativistic theory.
Most importantly, the huge data compression achieved via CD applies to both
the ERIs expressed in the AO basis and the ERIs expressed in the MO basis.
As a result, using CD, it is possible to bypass the disk-space and I/O bottleneck associated with MO based CC implementations,
as the MO transformed CVs can be typically kept in memory.
Moreover, the CD allows one to achieve a very efficient CC implementation that renders unprecedented calculations for large systems possible.\cite{Nottoli2023}

CD is also particularly well suited for the SFDC approximation,\cite{Banerjee2023}
as, based on the structure of the SFDC two-electron integrals, one can in fact expect that the number of CVs required to represent the ERIs up to a given accuracy
to be very similar to the number of vectors required in a non-relativistic calculation with the same AO basis.

In this contribution, we report on an implementation of CD for the SFDC two-electron integrals and use the resulting CVs to perform the SCF calculation and the integral transformation. We also interface our CD-SFDC implementation with an existing highly efficient non-relativistic CD-CC code developed by some of us.\cite{Nottoli2023}
Given that the CC step is completely identical in non-relativistic and SFDC calculations, and provided that the SCF and integral transformation steps with CD are not exceedingly more expensive than their non-relativistic counterparts, a whole CD-SFDC-CC calculation can be completed at (almost) non-relativistic cost.
We will give in this paper examples that show that this is indeed the case.

\section{Theory}\label{sec:theory}

Throughout this paper, we will use atomic units ($\hbar=e=m_e=1$)
and adopt the usual convention for indices:
Greek letters $\mu,\nu,\rho,\sigma,\ldots$ denote AOs (basis functions)
and Latin letters $p,q,r,s,\ldots$ refer to MOs.

\subsection{Spin-free Dirac-Coulomb approach}\label{sec:SFDC}

In the relativistic four-component schemes,
the DC Hamiltonian reads as
\cite{Dyall2007}
\begin{equation}\label{eq:HDC}
\HDC = \sum_i \hD(i) + \sum_{i<j}\gop(i,j) \,,
\end{equation}
where $i$ and $j$ run over the number of electrons.
$\hD(i)$ is the one-electron Dirac operator acting on $i$-th electron
\begin{equation}\label{eq:hD}
\hD(i) =
\begin{pmatrix}
    \V & c \vs\cdot\vp \\
    c \vs\cdot\vp & \V - 2c^2\\
\end{pmatrix}
\end{equation}
and $\gop(i,j)$ is the two-electron operator for the instantaneous electron-electron interaction
\begin{equation*}
\gop(i,j) = \frac{1}{r_{ij}} = \frac{1}{\abs{\Vec{r}_j-\Vec{r}_i}} \,.
\end{equation*}
In Eq.~\eqref{eq:hD}, $\vp$ is a momentum operator,
$\V$ is a one-particle operator for the Coulomb interaction of an electron with a nucleus,
$\vs=\left(\sx,\sy,\sz\right)$ the vector of the Pauli matrices,
and $c$ the speed of light.
The one-electron part of the DC Hamiltonian, a 2-by-2-block operator matrix,
acts on a four-component Dirac spinor $\ketpsi{p}$
(the relativistic counterparts to orbitals in the non-relativistic case)
that is composed of a large- (L) and a small-component (S) two-component spinor
($\alpha$ and $\beta$ denote spins)
\begin{equation*}
\ketpsi{p}=
\begin{pmatrix}
{\psiL_p} \\
{\psiS_p}
\end{pmatrix}
\,,
\qquad
\ketLS{p}=
\begin{pmatrix}
{\psi_p^{({\rm L,S}),\alpha}}\\
{\psi_p^{({\rm L,S}),\beta}}
\end{pmatrix}
\,.
\end{equation*}
Thus, in the full four-component schemes one faces the following
one- and two-electron matrix elements
\begin{equation}\label{eq:hD-matel}
\brapsi{p}\hD\ketpsi{q}
=
\braL{p}\V\ketL{q}
    + c \braS{p}\vs\cdot\vp\ketL{q}
    + c \braL{p}\vs\cdot\vp\ketS{q}
    + \braS{p}\left(\V - 2c^2\right)\ketS{q} \,,
\end{equation}
and
\begin{equation}\label{eq:2e-matel}
\twoelintpsi{p}{r}{q}{s}
= \LLLL{p}{r}{q}{s}
    +\LLSS{p}{r}{q}{s}
    +\SSLL{p}{r}{q}{s}
    +\SSSS{p}{r}{q}{s}
\,,
\end{equation}
respectively.
We use Mulliken notation for the two-electron repulsion integrals, i.e.,
\begin{equation*}
\twoelintpsi{p}{q}{r}{s}
= \int_{\mathcal{V}} \int_{\mathcal{V}} \psi_p^*(\Vec{r}_1)\psi_q(\Vec{r}_1)
            r_{12}^{-1}
            \psi_r^*(\Vec{r}_2)\psi_s(\Vec{r}_2)
            {\rm d} V_1 {\rm d} V_2 \,.
\end{equation*}

As originally noted by Kutzelnigg \cite{Kutzelnigg1984} 
and later developed by Dyall \cite{Dyall1994},
one can, motivated by the kinetic-balance condition,\cite{Stanton1984}
introduce a so-called pseudo-large component $\phiL$ to represent the small component $\psiS$
\begin{equation}\label{eq:KBC}
\ket{\psiS} = \frac{1}{2c}\vs\cdot\vp~\ket{\phiL}.
\end{equation}
The pseudo-large component $\phiL$
possesses the same symmetries as the large component $\psiL$
and, for electronic solutions, is of the same order of magnitude as the large component.
Therefore,
the pseudo-large component can be expanded in the same basis set as the large component,
in this way eliminating the need to specify a basis set for the small component.
Using the relation~\eqref{eq:KBC},
one obtains for the matrix elements of the DC Hamiltonian,
Eqs.~\eqref{eq:hD-matel} and \eqref{eq:2e-matel},
\begin{equation}\label{eq:hD-matel-KBC}
\brapsi{p}\hD\ketpsi{q}
=
\braL{p}\V\ketL{q}
    + \brapL{p}\T\ketL{q}
    + \braL{p}\T\ketpL{q}
    + \brapL{p}\left(\frac{1}{4c^2} \vs \cdot \vp \V \vs \cdot \vp - \T \right)\ketpL{q}
\end{equation}
and
\begin{equation}\label{eq:2e-matel-KBC}
\twoelintpsi{p}{r}{q}{s}
= \LLLL{p}{r}{q}{s}
+\frac{1}{4c^2}\LLPP{p}{r}{q}{s}
+
\end{equation}
\begin{equation*}
+\frac{1}{4c^2}\PPLL{p}{r}{q}{s}
+\frac{1}{16c^4}\PPPP{p}{r}{q}{s} \,,
\end{equation*}
respectively.

Next, using
the Dirac identity
(here $\vA$ and $\vB$ denote arbitrary vector operators)
\begin{equation}\label{eq:DI}
\vs\cdot\vA~\vs\cdot\vB = \vA\cdot\vB + i \vs\cdot\left(\vA\cross\vB\right) \,,
\end{equation}
one can rewrite the matrix elements of the DC Hamiltonian,
Eqs.~\eqref{eq:hD-matel-KBC} and \eqref{eq:2e-matel-KBC},
to a convenient form
where
the spin-independent (spin-free, SF)
and the spin-dependent (spin-orbit, SO) parts
are clearly separated.

In the case of the one-electron part, Eq.~\eqref{eq:hD-matel-KBC},
we apply the Dirac identity~\eqref{eq:DI} to the last term in Eq.~\eqref{eq:hD-matel-KBC} only
(all the other terms are already spin-independent).
We obtain
\begin{equation*}
\brapL{p}\left(\frac{1}{4c^2} \vs \cdot \vp \V \vs \cdot \vp - \T \right)\ketpL{q}
= \brapL{p}\left[\frac{1}{4c^2} \vp \V \cdot \vp + i \frac{1}{4c^2} \vs \cdot \left( \vp \V \cross \vp \right) - \T \right]\ketpL{q} \,,
\end{equation*}
where
\begin{equation*}
\brapL{p}\left(\frac{1}{4c^2}\vp\V\cdot\vp - \T\right)\ketpL{q}
\end{equation*}
is
the spin-independent
and
\begin{equation*}
i \frac{1}{4c^2} \brapL{p} \vs \cdot \left( \vp \V \cross \vp \right) \ketpL{q}
\end{equation*}
the spin-dependent part.
The SF and SO two-electron parts are separated similarly from Eq.~\eqref{eq:2e-matel-KBC};
a detailed derivation may be found elsewhere.

Finally,
due to the much smaller magnitude of the SO contributions,
we may neglect the SO terms and
thereby
obtain the so-called SFDC approximation.
The matrix elements of the SFDC Hamiltonian read as
\begin{equation}\label{eq:hDSF}
\brapsi{p}\hD\ketpsi{q}^{\rm SF}
=
\braL{p}\V\ketL{q}
+ c \brapL{p}\T\ketL{q}
+ c \braL{p}\T\ketpL{q}
+ \brapL{p}\left(\frac{1}{4c^2}\vp\V\cdot\vp - \T\right)\ketpL{q} \,,
\end{equation}
and
\begin{equation}\label{eq:GSF}
\twoelintpsi{p}{q}{r}{s}^{\rm SF}
=
\LLLL{p}{r}{q}{s}
+\frac{1}{4c^2}\LLSSSF{p}{r}{q}{s}
+\frac{1}{4c^2}\SSLLSF{p}{r}{q}{s}
+
\end{equation}
\begin{equation*}
+\frac{1}{16c^4}\SSSSSF{p}{r}{q}{s} \,.
\end{equation*}

As already mentioned, the large, $\psiL$, and pseudo-large, $\phiL$, components
can be expanded in the same AO basis set $\{\mu\}$
\begin{equation*}
\ketL{p} = \sum_{\mu} \CL{\mu p} \ket{\mu} \,,\quad
\ketpL{p} = \sum_{\mu} \CS{\mu p} \ket{\mu} \,.
\end{equation*}
Thus, for a SFDC calculation we need to evaluate the following matrix elements,
cf.~Eqs.~\eqref{eq:hDSF} and \eqref{eq:GSF},
\begin{equation}\label{eq:hDSF-AO}
\DLL{\mu\nu} = \bra{\mu}\V\ket{\nu} \,,\quad
\DLS{\mu\nu} = \DSL{\mu\nu} = \bra{\mu}\T\ket{\nu} \,,\quad
\DSS{\mu\nu} = \frac{1}{4c^2} \bra{\mu}\vp \V \cdot \vp \ket{\nu} - \bra{\mu}\T\ket{\nu}\,,
\end{equation}
and
\begin{equation}\label{eq:GSF-AO}
\WLL{\mu\nu,\rho\sigma} = \twoelint{\mu}{\nu}{\rho}{\sigma} \,,\quad
\WLS{\mu\nu,\rho\sigma} = \frac{1}{4c^2} \LLSSSFAO{\mu}{\nu}{\rho}{\sigma} \,,
\end{equation}
\begin{equation*}
\WSL{\mu\nu,\rho\sigma} = \frac{1}{4c^2} \SSLLSFAO{\mu}{\nu}{\rho}{\sigma} \,,\quad
\WSS{\mu\nu,\rho\sigma} = \frac{1}{16c^4} \SSSSSFAO{\mu}{\nu}{\rho}{\sigma} \,.
\end{equation*}

The evaluation of the one-electron integrals \eqref{eq:hDSF-AO} is relatively cheap,
whereas the two-electron integrals \eqref{eq:GSF-AO} are very costly.
Note
that in comparison with the non-relativistic case,
there are now four-times as many ERIs.
In addition, the ERIs involving the small component require the calculation of derivative integrals,
which increases their price.
For brevity and clarity,
we will sometimes use the common short-hand forms to refer to the above SF two-electron integrals:
$\intLLLL$, $\intLLSS$, $\intSSLL$, $\intSSSS$, respectively.

\subsection{Cholesky decomposition}\label{sec:CD}

CD is a well-known mathematical tool that allows one to
decompose a symmetric positive-definite matrix into a triangular matrix of so-called CVs.
As originally suggested by Beebe and Linderberg,\cite{Beebe1977}
this method may be applied to decompose the positive-semidefinite ERI matrix:
\begin{equation*}   
W_{\mu\nu,\rho\sigma} = \twoelint{\mu}{\nu}{\rho}{\sigma} \approx \sum_{P=1}^{\ncv} L_{\mu\nu}^{P} L_{\rho\sigma}^{P} \,,
\end{equation*}
where $L_{\mu\nu}^{P}$ is the $P$-th Cholesky vector.

The general algorithm for a CD of an ERI matrix goes as follows:\cite{Beebe1977,Koch2003}
\begin{enumerate}
\item Compute all diagonal elements $\widetilde{W}_{\mu\nu,\mu\nu}=W_{\mu\nu,\mu\nu}$.
\item Find the largest (updated) diagonal element $\widetilde{W}_{\mu\nu,\mu\nu}$.
\item Get the corresponding $P$-th CV:\\
\begin{equation*}
L_{\rho\sigma}^{P}
=
\left( \widetilde{W}_{\mu\nu,\mu\nu} \right)^{-1/2}
\left(
    W_{\rho\sigma,\mu\nu} - \sum_{R=1}^{P-1} L_{\rho\sigma}^{R} L_{\mu\nu}^{R}
\right) \,.
\end{equation*}
\item Update the vector of diagonal elements
\begin{equation*}
\widetilde{W}_{\mu\nu,\mu\nu} = W_{\mu\nu,\mu\nu} - \sum_{R=1}^{P} L_{\mu\nu}^{R} L_{\mu\nu}^{R} \,.
\end{equation*}
\item Continue to step 2. until the largest diagonal element is smaller than a given Cholesky threshold $\cdthr$.
\end{enumerate}
The error of this decomposition is controlled rigorously via the Cauchy-Schwarz inequality,\cite{Koch2003}
which relates the error of any matrix element to the error in the diagonal elements,
\begin{equation*}
\abs{ W_{\mu\nu,\rho\sigma} - \sum_P L_{\mu\nu}^{P} L_{\rho\sigma}^{P} }
\le
\abs{ W_{\mu\nu,\mu\nu} - \sum_P L_{\mu\nu}^{P} L_{\mu\nu}^{P} }^{1/2}
\abs{ W_{\rho\sigma,\rho\sigma} - \sum_P L_{\rho\sigma}^{P} L_{\rho\sigma}^{P} }^{1/2}
\le
\cdthr \,.
\end{equation*}

In the case of relativistic SFDC calculations,
there are four types of two-electron integrals, see~Eq.~\eqref{eq:GSF} and expressions~\eqref{eq:GSF-AO};
hence the dimension of the two-electron matrix $\mW$ is twice that of the non-relativistic one:
\begin{equation*}
\mW
=
\begin{pmatrix}
    \mWLL & \mWLS \\
    \mWSL & \mWSS
\end{pmatrix}
\end{equation*}
and the CVs are now composed of a large- and a small-component part:
\begin{equation*}
\mL^{P}
=
\begin{pmatrix}
    \mLLP{P} \\
    \mLSP{P}
\end{pmatrix} \,.
\end{equation*}
One could, of course, straightforwardly apply the CD procedure to the whole matrix $\mW$.
However, there is a more efficient approach, as also suggested for the two-step algorithm in Ref.~\citenum{Banerjee2023}:
As the small-component integrals $W_{\mu\nu,\rho\sigma}^{\rm SS}$
are suppressed by a factor of $1/(16c^4)$, cf. Eq.~\eqref{eq:GSF-AO},
their contribution is much smaller than that of the large-component integrals $W_{\mu\nu,\rho\sigma}^{\rm LL}$.
Therefore it suffices to consider for pivoting\cite{Koch2003} the large-component diagonals $W_{\mu\nu,\mu\nu}^{\rm LL}$ only.
The significantly more expensive small-component integrals $W_{\mu\nu,\rho\sigma}^{\rm SS}$,
which then do not have to be calculated at all now,
are reconstructed in the calculations from the small-component part of CVs to sufficient accuracy.
Specifically,
the above outlined general CD algorithm
is modified as follows:
we now work with the large-component diagonals only, i.e.,
we substitute everywhere
$\widetilde{W}_{\mu\nu,\mu\nu}=\widetilde{W}_{\mu\nu,\mu\nu}^{\rm LL}$,
and we now need to compute both the large- and small-component parts
of the CVs (step 3. above), i.e.,
\begin{equation*}
L_{\rho\sigma}^{{\rm L},P}
=
\left( \widetilde{W}_{\mu\nu,\mu\nu}^{\rm LL} \right)^{-1/2}
\left(
    W_{\rho\sigma,\mu\nu}^{\rm LL} - \sum_{R=1}^{P-1} L_{\rho\sigma}^{{\rm L},R} L_{\mu\nu}^{{\rm L},R}
\right)
\end{equation*}
and
\begin{equation*}
L_{\rho\sigma}^{{\rm S},P}
=
\left( \widetilde{W}_{\mu\nu,\mu\nu}^{\rm LL} \right)^{-1/2}
\left(
    W_{\rho\sigma,\mu\nu}^{\rm SL} - \sum_{R=1}^{P-1} L_{\rho\sigma}^{{\rm S},R} L_{\mu\nu}^{{\rm L},R}
\right) \,.
\end{equation*}
Otherwise, the algorithm remains unchanged.

The same idea can be also applied to the two-step CD algorithm,\cite{Folkestad2019}
as recently used in Ref.~\citenum{Banerjee2023}.
In the two-step scheme, which is computationally more efficient than the original one-step approach,
one first determines only the Cholesky basis using the procedure described in this section, but discarding all the elements of the computed CVs corresponding to pairs of indices that correspond to updated diagonal elements smaller than the Cholesky threshold. Then, once the Cholesky basis has been determined, one computes the CVs using the CD -- density-fitting equivalence.\cite{Aquilante2008a} In other words, one uses the Cholesky basis, denoted with indices $P$, as a density-fitting basis 
\begin{equation}
    \label{eq:CDDF}
    (\mu\nu|\rho\sigma) = \sum_{PQ}(\mu\nu|P)(P|Q)^{-1}(Q|\rho\sigma)
\end{equation}
and then computes the CVs as
\begin{equation}
    \label{eq:secondstep}
    L^P_{\mu\nu} = \sum_Q(\mu\nu|Q)K^{-T}_{QP},
\end{equation}
where $K^{-T}$ is the inverse transpose Cholesky factor of the metric $(P|Q)$, i.e., $(P|Q) = KK^T$.
In Ref.~\citenum{Banerjee2023} it was pointed out that
while the user-defined Cholesky threshold $\cdthr$ provides an upper bound to the error
of the CD of the large-component ERIs,
it does not bound the error of the small-component ERIs;
hence a variational collapse might be in principle possible in such four-component calculations.

\subsection{AO $\rightarrow$ MO transformation of Cholesky vectors}
\label{sec:AO2MO}

In general, the AO$\rightarrow$MO transformation of CVs
is achieved
via the transformation of the CVs
using the MO coefficients
obtained from the HF calculation,
see the first part of Eq.~\eqref{eq:AO2MO}.
In SFDC calculations,
the CVs comprise two parts:
a large- and a small-component part.
Hence the double summation in the AO$\rightarrow$MO transformation
splits in fact into two contributions:
\begin{equation}\label{eq:AO2MO}
L_{pq}^P = \sum_{\mu,\nu} L_{\mu\nu}^P C_{\mu p} C_{\nu q}
=
  \sum_{\mu,\nu}^{\rm L} L_{\mu\nu}^{{\rm L},P} C_{\mu p}^{\rm L} C_{\nu q}^{\rm L}
+ \sum_{\mu,\nu}^{\rm S} L_{\mu\nu}^{{\rm S},P} C_{\mu p}^{\rm S} C_{\nu q}^{\rm S} \,.
\end{equation}
Within the no-pair approximation in relativistic calculations,
the positronic solutions from the HF step are neglected in the subsequent CC treatment,
i.e.,
the above MO indices $p$, $q$ only run over electronic states.
Thus, 
the structure of CVs in the MO basis is exactly the same
for non-relativistic and SFDC calculations;
hence non-relativistic CD-CC codes can be easily used.

\subsection{SFDC-CCSD with Cholesky vectors}
\label{sec:CCSD}

As mentioned in the Introduction,
the CD-SFDC-CC step
is exactly the same as its non-relativistic CD-CC counterpart.
We use a recently reported highly-effective CD-CC implementation,\cite{Nottoli2023}
which fully exploits Abelian point-group symmetry and the use of CD of ERIs.

\section{Results \& Discussion}\label{sec:results}
In this section, we present various examples to demonstrate the accuracy and efficiency of our CD based SFDC implementation.
First, we discuss the effects of restricting the Cholesky pivots to the large-components elements, as suggested in section \ref{sec:CD},
and then we compare the accuracy of CD based calculations with those of their traditional counterparts.
Then, we analyze the compression introduced by adopting the CD and compare it with the non-relativistic case,
showing how using the CD in a SFDC calculation is particularly advantageous,
and illustrate how this makes integral transformations rather inexpensive even for large systems.
Finally, we focus on the performance of our CD-SFDC-CC implementation and demonstrate that,
thanks to the CD, any calculation that can be done in a non-relativistic framework can also be performed using a four-component SFDC relativistic treatment.
In particular, we perform a series of calculations
for the closed-shell \ch{X(CO)4}, X=Ni,Pd,Pt, molecules, for which scalar-relativistic effects play an important role.
All calculations for which we report timings information
were run on a single-cluster Intel\textsuperscript{\textregistered} Xeon\textsuperscript{\textregistered} Gold 6342 node
equipped with
48 CPUs
(of which 1, 8 or 16 were used,
as specified in the pertinent results)
running at
its maximum single-core frequency
3.50
GHz.

\subsection{Implementation details}\label{sec:implementation}

The above described strategy for the CD of relativistic SFDC ERIs
has been implemented within a development version of the \cfour program package \cite{cfour}
to allow for high-accuracy cost-effective scalar-relativistic calculations
up to the CCSD level.
The one- and two-electron integrals are calculated using the McMurchie-Davidson scheme.\cite{McMurchie1978,Stopkowicz2012}
The $\intLLSS$ and $\intSSLL$ integrals are in practice evaluated 
as linear combinations $\intLLSS~\pm~\intSSLL$ to preserve the symmetry
upon electron exchange $1 \leftrightarrow 2$.
The CD of the ERIs for SFDC calculations
has been implemented both in its one- and two-step variants
with the possibility to choose whether
the entire diagonal should be considered
or only the large-component part (hereafter referenced as option `FULL' and `LARGE', respectively).
Both CD algorithms have been also parallelized using OpenMP.
The existing non-relativistic SCF procedure has been extended to also allow the SFDC case.
The transformation of relativistic CVs from AO to MO basis has been added as well
and an interface to a recently reported highly efficient CD-CC program
\cite{Nottoli2023}
has been provided.
In addition, the implementation
is available
both with and without exploiting Abelian point-group symmetry of the given system.

\subsection{Accuracy}
\begin{figure}[h!]
    \center
    \includegraphics[width=0.9\textwidth]{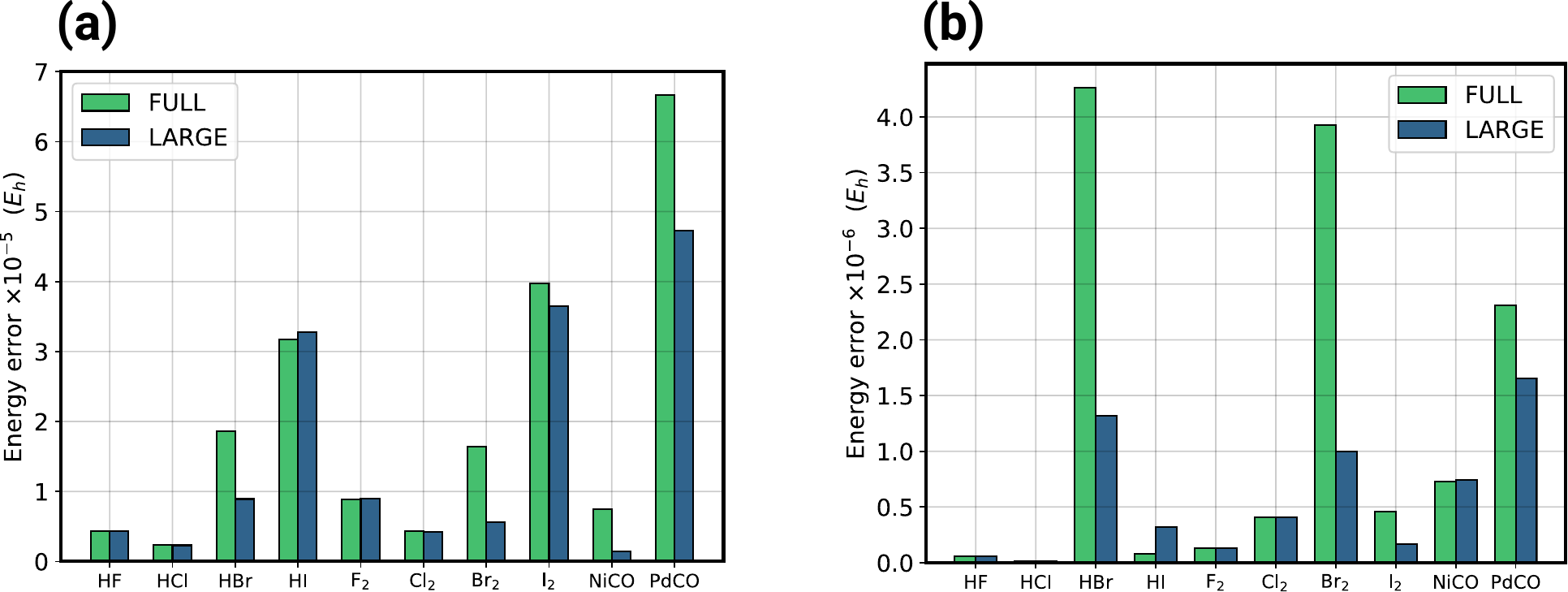}
    \caption{Error on the SCF (panel a) and correlation (panel b) energies, computed as in Eq.~\eqref{eq:errorene_cd_vs_std}, for CD-SFDC-CCSD calculations performed using the unc-ANO-RCC basis set, using both the FULL Cholesky basis and the LARGE Cholesky basis. The Cholesky tolerance is set to $10^{-5}$.}
    \label{fig:errenergy}
\end{figure}

To assess the accuracy of the Cholesky basis obtained by only considering the large-component pivots, we compare SCF energies and CCSD correlation energies for a range of small molecules, computed with the LARGE and the FULL Cholesky bases (the corresponding energy is denoted as $E_{\rm CD}$), to the energy computed without CD (denoted as $E_{\rm STD}$), and report the associated error
\begin{equation}
    {\Delta E_{\rm Error}} = |E_{\rm STD} - E_{\rm CD}|.
    \label{eq:errorene_cd_vs_std}
\end{equation}
While CD affords a rigorous \textit{a priori} error estimate only for the two-electron integrals, we expect the error in the energies to be of the same order of magnitude as the threshold used to terminate the decomposition. All the calculations were performed using the uncontracted ANO-RCC (unc-ANO-RCC) basis set\cite{roos2004relativistic} and a Cholesky threshold $\cdthr = 10^{-5}$. The results are displayed in Figure~\ref{fig:errenergy}.
For all tested systems, the error in the SCF energy (left panel) is indeed $\mathcal{O}(10^{-5})$, independent of whether the FULL or the LARGE Cholesky basis is used. This confirms that the CD of the ERIs is an accurate approximation also in the scalar-relativistic case, and that using the LARGE basis does not introduce a significant error in the energy. This is particularly relevant, as computing the CD using the LARGE basis does not require one to evaluate any of the expensive $\intSSSS$ integrals, which further improves the overall efficiency of the implementation.
Another consequence of the small magnitude of the $\intSSSS$ integrals can be seen in the number of CV; in fact, the majority of them stem exclusively from the large-component pivots. Therefore, as reported in Table~\ref{tab:cvfull_vs_cvlarge}, the difference between $\ncv^{\rm FULL}$ and $\ncv^{\rm LARGE}$ is rather small.
However, the LARGE basis is only approximately a subset of the FULL one since the consideration of the small pivots (slightly) affects the diagonal update, hence, the determination of the large pivots. Nonetheless, we can conclude that the LARGE basis has enough flexibility to afford an accurate representation of the product densities, as expected given the small magnitude of the $\intSSSS$ integrals.
\begin{table}[h!]
    \centering
    \begin{tabular}{ccc}
       \toprule
        Molecule & $\ncv^{\rm FULL}$ & $\ncv^{\rm LARGE}$ \\
        \midrule
        \ch{HF}  & 799 & 797 \\
        \ch{HCl} & 905 & 898 \\
        \ch{HBr} & 1212 & 1178 \\
        \ch{HI}  & 1470 &  1417 \\
        \midrule
        \ch{F2}  & 1161 &  1157 \\
        \ch{Cl2} & 1340 &  1326 \\
        \ch{Br2} &  1238 &  1206 \\
        \ch{I2}  &  1480 & 1424 \\
        \midrule
        \ch{NiCO} & 2316 & 2296 \\
        \ch{PdCO} & 2475 & 2432 \\
        \bottomrule
    \end{tabular}
    \caption{Comparison between the number of Cholesky vectors $\ncv$ when using the FULL Cholesky basis and the LARGE Cholesky basis. The unc-ANO-RCC basis set was used with a Cholesky threshold of $10^{-5}$.}
    \label{tab:cvfull_vs_cvlarge}
\end{table}
The error in the correlation energies (Figure~\ref{fig:errenergy}, right panel) is one order of magnitude smaller than for its SCF counterpart for both the large and full component. This behavior is consistent with what has been found in a recent publication by some of us\cite{Zhang2024} and further demonstrates that accurate correlation energies can be computed using the CD. In conclusion, our numerical tests document the accuracy of the CD for the matrix elements of the SFDC Hamiltonian and that the LARGE Cholesky basis, obtained by dropping all small-component pivots during the CD procedure, provides a very good approximation to the FULL basis. This is remarkable, as the LARGE Cholesky basis does not require one to compute any $\intSSSS$ two-electron integral, which not only is computationally advantageous, as such integrals are particularly expensive to evaluate, but also dramatically simplifies the implementation.

\subsection{Efficiency} 

\begin{figure}[htbp]
     \center
     \includegraphics[scale=0.8]{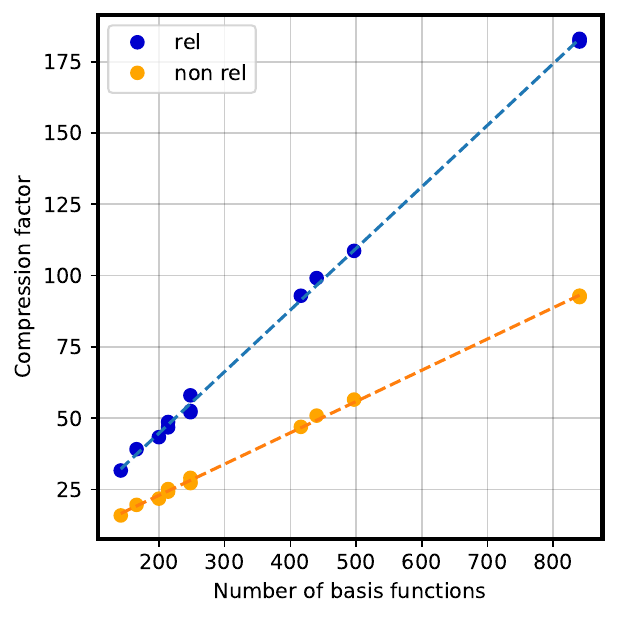}
     \caption{Comparison between the SFDC and the non-relativistic compression factors; the Cholesky tolerance in the calculation was $10^{-4}$ and the calculation were performed using the FULL Cholesky basis together with an unc-ANO-RCC basis set. The lines were obtained by fitting the data to a linear function. We obtained y =  0.109 x + 1.221 for the non-relativistic case ($R^2=0.99$) and y =  0.215 x + 1.730 for the relativistic one ($R^2=0.99$)}
      \label{fig:compression_factor}
\end{figure}

In this section,
we demonstrate that the CD affords a particularly compact representation of the SFDC two-electron integrals by reporting the compression factor;
\textit{i.e.}, the ratio between the number of CVs required for an exact (\textit{i.e.}, untruncated) decomposition
and the number of vectors  ($\ncv$) obtained by truncating the decomposition at a given threshold.
We compare the non-relativistic and SFDC compression factors that are defined as follows:
\begin{eqnarray}
    f_{\rm rel} & = & \frac{N(N+1)}{\ncv} \\
    f_{\rm nonrel} & = & \frac{N(N+1)/2}{\ncv} \,,
    \label{eq:comp_fac}
\end{eqnarray}
where $N$ is the number of basis functions. Note that in the SFDC case,
the exact number of pivots is twice of its non-relativistic counterpart
due to the presence of the small-component integrals.
In Figure~\ref{fig:compression_factor}, we report the relativistic and non-relativistic compression factors for a set of molecules (\ch{HF}, \ch{HCl}, \ch{HBr}, \ch{HI}, \ch{F2}, \ch{Cl2}, \ch{Br2}, \ch{I2}, \ch{NiCO}, \ch{PdCO}, \ch{PtCO}, \ch{TcF6}, \ch{RuF6}, \ch{RhF6}), using the unc-ANO-RCC basis set. To have a more consistent comparison between the SFDC and the non-relativistic factors, we use the FULL Cholesky basis. For all calculations, the threshold used is $\cdthr = 10^{-4}$. 

The SFDC compression factors are about twice as large as their non-relativistic counterparts,
and
they both scale linearly with the system size.
This is to be expected given
that the SFDC and non-relativistic Cholesky bases consists of more or less the same number of vectors
and
the number of CVs is $\mathcal{O}(N)$.
Thus, given that the length of a CV is $\mathcal{O}(N^2)$,
the memory required to store the CVs scales as $\mathcal{O}(N^3)$.
The crucial aspect here is that,
contrary to what is observed for the standard two-electron integrals,
this also applies to the CVs in the MO basis.
In other words, 
the CD allows one to store the MO transformed two-electron integrals in memory even for large molecules.
Consequently,
the transition from the AO to the MO representation can be performed very efficiently without any disk I/O impediment.
In addition,
the AO to MO transformation of the CVs 
consists of only
matrix-matrix multiplications,
see Eq.~\eqref{eq:AO2MO},
which can be computed at $\mathcal{O}(N^4)$ cost and very efficiently.
Although
such integral transformations are twice as expensive for SFDC calculations
as for their non-relativistic counterparts,
as shown in Eq.~\eqref{eq:AO2MO},
the increase in the overall computational cost is negligible.

To demonstrate the efficiency of the CD in electron-correlated SFDC calculations, we compute the MP2 energy of a very large molecule, auranofin (C$_{20}$H$_{34}$AuO$_9$PS), a molecule that has been investigated for its anti-cancer activity,\cite{mirabelli1985evaluation,marzano2007inhibition} and whose geometry has been taken from Ref.~\citenum{dos2014reactivity}.

\begin{figure}[h!]
    \center
   \includegraphics[scale=0.1]{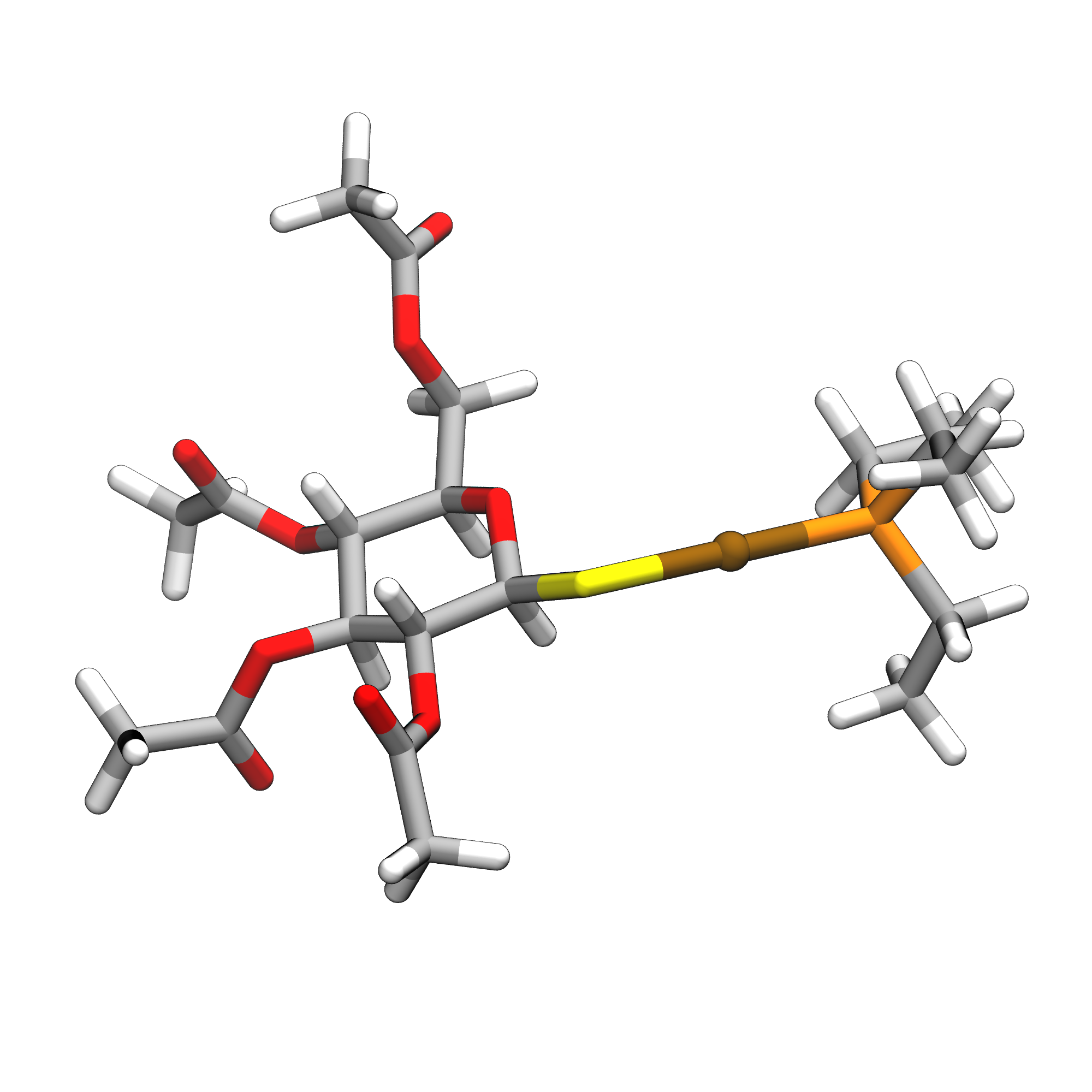}
    \caption{Structure of the auranofin (C$_{20}$H$_{34}$AuO$_9$PS) molecule. Carbon is represented in silver, oxygen in red, hydrogen in white, sulfur in yellow, phosphorous in orange, and gold in brown.}
    \label{fig:auranofin}
\end{figure}
The CD-SFDC-MP2 calculation was performed using the uncontracted TZP basis set (1544 basis functions per component),\cite{barbieri2006gaussian} using both the FULL and the LARGE Cholesky bases and a Cholesky threshold $\cdthr=10^{-4}$. We also performed a non-relativistic CD-MP2 calculation as a reference.

\begin{table}[h!]
    \centering
    \begin{tabular}{cccccc}
       \toprule
        SFDC & Cholesky basis ($\mathcal{B}$) & Energy ($E_{\rm h}$) & t$_{\rm SCF}$ (it.) & t$_{\rm AO-MO}$ & t$_{\bra{ij}\ket{ab}}$\\
        \midrule
          ON  & FULL   & -21221.13008  &  86.8 (45)  & 1.40 & 10.1 \\
          ON  & LARGE  & -21221.12961  &   83.0 (45) & 1.15 & 9.96 \\
          OFF & --     & -20064.41211  &  17.7 (44)  & 0.49 & 9.52 \\
        \bottomrule
    \end{tabular}
    \caption{Results obtained from CD-MP2 calculations, with and without relativity, for auranofin (C$_{20}$H$_{34}$Au$_9$PS) with an uncontracted TZP basis set (1544 basis functions) using a CD threshold equal to $10^{-4}$. We report the total (MP2) energy in Hartree ($E_{\rm h}$), the total time required to converge the SCF (t$_{\rm SCF}$) and in parentheses the number of iterations, the time required to transform the CVs from the AO to the MO representation (t$_{\rm AO-MO}$), and the time needed to build $\bra{ab}\ket{ij}$ integrals. Energies in Hartree and timings are in minutes.}
    \label{tab:auranofin}
\end{table} 

In Table~\ref{tab:auranofin}, we report the total MP2 energy and the timings for the most important steps of the calculations, namely, the solution to the SCF equations, the transformation of the CVs into the MO basis, and the generation of the $\langle ab|ij\rangle$ integrals required to compute the MP2 amplitudes and energies. For all calculations we used 16 OpenMP threads. It is noteworthy that the cost of the calculation is dominated by the SCF step, which is about four times more expensive in the SFDC case, as expected. 
The integral transformation step is overall particularly cheap, as it only requires $\mathcal{O}(ON^2\ncv + OVN\ncv)$ floating-point operations, that are performed efficiently using matrix-matrix multiplications distributed among the available cores, 
and as expected, the SFDC transformation of the CVs is about two times more expensive as their non-relativistic counterpart. The slight overhead in the SFDC case is probably due to a small difference in the $\ncv$ among the three cases. Finally, the generation of the integrals requires a $\mathcal{O}(O^2V^2\ncv)$ matrix-matrix multiplication plus a transposition, and dominates the MP2 step of the calculation. 

\subsection{"Non-relativistic cost" of CD-SFDC-CC calculations}\label{sec:SFDC-CD-CC}

In this section,
we discuss calculations illustrating the main point of this paper:
CD-SFDC-CC calculations may be run at the cost of corresponding non-relativistic calculations.
We choose as an example
a series of the tetrahedral \ch{X(CO)4}, X=Ni,Pd,Pt, molecules.
The results are summarized in Tables~\ref{tab:NiCO4-walltime}--\ref{tab:PtCO4}.

All calculations
were performed with the \cfour program package
using an unc-ANO-RCC basis set;\cite{roos2004,roos2005}
this produces
42 doubly occupied and 974 virtual orbitals
(1016 AOs in total) for \ch{Ni(CO)4},
51 doubly occupied and 989 virtual orbitals
(1040 AOs in total) for \ch{Pd(CO)4},
and
67 doubly occupied and 1008 virtual orbitals
(1075 AOs in total) for \ch{Pt(CO)4}.
The frozen-core approximation was employed:
$1s$ for carbon and oxygen,
$1s2s2p$ for nickel,
$1s2s2p3s3p$ for palladium,
and $1s2s2p3s3p3d4s4p$ for platinum
leading to 13, 17, and 26, respectively, frozen orbitals.
We also note that
damping during the iterative SCF solution
and DIIS acceleration were necessary to achieve convergence in the SCF step.
In addition,
an \textit{a priori} specification of the occupation vector
further boosted the SCF convergence
as otherwise the occupation was often changing within the first few iterations.
We used the experimentally determined structure for \ch{Ni(CO)4} \cite{Hedberg1979}
and previously published theoretical structures for \ch{Pd(CO)4} and \ch{Pt(CO)4} molecules. \cite{Frenking2021}
The \ch{Ni(CO)4} calculations used for obtaining timings were run on only one core
to allow for a fair comparison with unparallelized CCSD codes
without the CD.
The SCF solution usually converged (threshold set to $10^{-7}$ for the maximum change in the density matrix) in
31 steps for \ch{Ni(CO)4},
35 steps for \ch{Pd(CO)4}, and
39 steps for \ch{Pt(CO)4}.
The CCSD step (threshold set to $10^{-7}$ for the maximum norm of the residual)
usually converged in approximately 20 steps.
In timing comparisons (\textit{e.g.}, non-relativistic vs. SFDC),
the number of iterations was always the same.

As already mentioned in the Introduction,
the major obstacle in using MO based CC codes,
which would reduce the cost of SFDC-CC calculations to those of the non-relativistic case,
is the enormous memory requirement
due to the $\mathcal{O}(V^4)$ $\twoelint{a}{b}{c}{d}$ integrals.
For example, the above specified \ch{X(CO)4}, X=Ni,Pd,Pt,
CCSD calculations
would require 8--9 TB of memory only to store all unique two-electron integrals in the MO basis.
This difficulty is, however, completely overcome by using the CD,
as memory required to store the CVs scales as $\mathcal{O}(N^3)$
for the CVs expressed both in the AO and in the MO basis,
see the earlier discussion on the efficiency of the CD.
For example, for the \ch{Ni(CO)4} molecule,
we need mere 7--14 GB to store the MO transformed CVs, see Table~\ref{tab:NiCO4-CD-memory}.
The whole CCSD calculation then requires 80--110 GB for single-thread
or 160--180 GB for 8-thread calculations, see Tables~\ref{tab:NiCO4-walltime-8cores} and \ref{tab:NiCO4-CD-memory} ;
cf. also the results for \ch{Pd(CO)4} and \ch{Pt(CO)4} in Tables~\ref{tab:PdCO4} and \ref{tab:PtCO4}.
Note that
the highest contribution to the CCSD memory strain (about half of the peak core memory) comes
from an intermediate term that scales as $\mathcal{O}(V^3)$.
Given that the SCF calculation with CD requires approximately
5--9 GB and 10--18 GB in the non-relativistic and SFDC case, respectively,
the memory requirements to run the whole CC calculation are clearly dictated by the CC step.

The total wall times of the CC calculations are driven by the CD-CC step, too,
see Tables~\ref{tab:NiCO4-walltime}, \ref{tab:NiCO4-walltime-8cores}, \ref{tab:PdCO4}, and \ref{tab:PtCO4}.
This part of the calculation amounts to approximately 80--90 \% of the total wall time in the SFDC-CC calculations
and approximately 95 \% of the total wall time in the non-relativistic case.
The remaining 10--20\% (5\%) contribution to the total wall time is due to
the three pre-CCSD steps:
the evaluation of integrals,
which is responsible for the vast majority of the 10--20 \% (or 5 \%, resp.),
the SCF calculation,
and the AO-MO transformation,
both of which require relatively very little time.
Here, we want to note that
the \ch{X(CO)4}, X=Ni,Pd,Pt, calculations
currently use the one-step CD algorithm,
as the Abelian point-group symmetry
is not fully implemented in the two-step algorithm yet.
As the two-step algorithm is
computationally substantially more efficient
than the one-step algorithm,
we can expect the contributions to the total wall time
to shift to an even more favorable ratio,
i.e.,
the integral evaluation step will consume
significantly less than the current 10--20\% (5\%) of total wall time,
and hence the dominance of the CC step will be complete.

Note also the huge cost reduction in comparison with calculations without CD,
see the results for the \ch{Ni(CO)4} molecule in Table~\ref{tab:NiCO4-walltime}.
The SCF and mainly the integral transformation steps are even several orders of magnitude slower without the CD.
Note especially in Table~\ref{tab:NiCO4-walltime}
that the AO-MO transformation step without the CD in the SFDC calculations requires
at least the same time as the whole CCSD step with CD.
Unfortunately,
the memory requirements for the CCSD  calculation for \ch{Ni(CO)4}
without CD
are enormous, as discussed above,
and thus reference calculations completely analogous to those with CD were not feasible (a corresponding SFDC scheme using AO algorithms, though possible but not very efficient in terms of computation times, has not been available to us).
For the non-relativistic CCSD calculation without CD for \ch{Ni(CO)4},
AO based algorithms were used to render the computation feasible by reducing the memory requirements, see Table~\ref{tab:NiCO4-walltime}.

Finally, we would like to show that
one can use the CD-SFDC-CCSD scheme for numerical geometry optimizations, for instance,
as we illustrate on the \ch{Ni(CO)4} molecule.
The optimized structures,
as well as the experimental ones \cite{Hedberg1979}
and other theoretical ones \cite{Frenking2021,McKinlay2015} for comparison,
are summarized in Table~\ref{tab:NiCO4-gopt}.
On the technical side, we note that
while geometry optimizations using Cholesky thresholds of $10^{-5}$ and $10^{-6}$
converged without problems,
the structure
determined with a Cholesky threshold of $10^{-4}$
had problems to converge to more than four decimal points
(denoted by asterisks in Table~\ref{tab:NiCO4-gopt}).
The scalar-relativistic value for the C-O bond length does not differ from the non-relativistic one
until the fourth decimal place,
while the value for Ni-C bond length already differs at the second decimal place.
This is to be expected:
carbon and oxygen are light elements, hence scalar-relativistic effects only contribute subtly,
whereas nickel is a heavy d-element, so that scalar-relativistic effects are much more pronounced.
A non-negligible contribution to the molecular structure,
namely bond elongation,
would come from the inclusion of triples (e.g. at the CCSD(T) level),
as one can see, e.g., from C-O bond length in the carbon-monoxide molecule:
$r=1.125~\text{\AA}$ and $r=1.132~\text{\AA}$ at the CCSD and CCSD(T) level, respectively
(non-relativistic calculation and unc-ANO-RCC basis set).
The CCSD(T) extension to the current CCSD method with Cholesky decomposition is currently under development.

\section*{Conclusions}

This work presents an efficient implementation of the CD of two-electron integrals within the SFDC approach
that reduces the cost of CD-SFDC-CC calculations
to the cost of the corresponding non-relativistic calculations.

In the integral evaluation step,
it suffices to consider only the ERIs between large components, \intLLLL,
for pivoting during the CD.
The thus obtained Cholesky basis has enough flexibility
to account for the small-component integrals, \intSSSS,
and we avoid the evaluation of these very expensive \intSSSS~ integrals.
We demonstrate this point in a series of SCF and CCSD energy calculations
and show that the error in the SCF or CCSD energy, respectively,
is indeed driven by the \textit{a priori} chosen Cholesky threshold
irrespective of whether the whole or only the large-component diagonal is considered.

In the subsequent MO based CC calculations,
the use of the CD
allows one to completely overcome
problems
associated with the enormous memory requirements
due to the necessary storage of the $\mathcal{O}(V^4)$ $\twoelint{a}{b}{c}{d}$ integrals.
The memory for the storage of CVs scales favorably as $\mathcal{O}(N^2\ncv)$
for CVs both in the AO and MO bases,
and therefore the entire AO-MO transformation of two-electron integrals,
which is itself very efficient as it involves only matrix-matrix multiplications,
can be easily completed only in memory.
Although for the SFDC AO-MO transformation of the the CVs,
there is a factor $2$ with respect to the non-relativistic case,
this transformation is far from being the bottleneck of the whole calculation,
and thus this increase in cost does not affect significantly the overall wall time.
We illustrate the very low cost of the AO-MO transformation
in CD-SFDC-CCSD calculations for \ch{Ni(CO)4}.
The memory required to store all unique two-electron integrals in the MO basis
would require almost 8 TB,
whereas mere 7--14 GB suffice for the CVs,
and
the AO-MO transformation step contributes to the total wall time with at most 0.1 \%.

Given the low cost and simplicity of the SFDC AO-MO transformation of the CVs,
one can
easily interface a CD-SFDC-SCF code
to already existing highly-efficient MO based non-relativistic CD-CC codes,
and thus be able to run SFDC-CC calculations even for large systems.
We present a series of illustrative CCSD energy calculations
for the highly symmetric tetrahedral \ch{X(CO)4}, X=Ni,Pd,Pt, molecules,
for which scalar-relativistic effects play an important role.
We show that
about 80--90 \% of the total wall time
as well as the total memory requirements of the CD-SFDC-CCSD calculations
are due to the CD-CCSD step,
which is identical for both the non-relativistic and the SFDC calculation.
Once the point-group symmetry is fully implemented within the two-step Cholesky algorithm,
the dominance of the CC step will be even much greater.
Therefore,
an increased cost on the SCF and AO-MO transformation level in a scalar-relativistic calculation
does not introduce a significant overhead to the whole calculation;
CD-SFDC-CC results are obtainable at approximately the same cost
as their non-relativistic CD CC counterparts.

Finally,
we would like to note that
a further huge cost reduction is achieved
via
the inclusion of Abelian point-group symmetry
and via efficient parallelization using OpenMP.
Thus,
the combination of all the herein described improvements
allows one to complete CD-SFDC-CC calculations
that would be otherwise impossible.


\begin{acknowledgement}
This paper is dedicated to Rodney J. Bartlett as a recognition of his many important contributions to the field of coupled-cluster theory. In addition, JG thanks him for his friendship and for his constant support since postdoctoral stay in his group.
T.N. and F.L. acknowledge financial support from ICSC-Centro Nazionale di Ricerca in High Performance Computing, Big Data, and Quantum Computing, funded by the European Union -- Next Generation EU -- PNRR, Missione 4 Componente 2 Investimento 1.4., and from the Italian Ministry of Research under grant number 2022WZ8LME$\_$002, while the work in Mainz was funded funding by the Deutsche Forschungsgemeinschaft (DFG) within
the project B5 of the TRR 146 (Project No. 233630050).


\end{acknowledgement}






\bibliography{references}

\begin{sidewaystable}[htb!]
    \centering
    \caption{Non-relativistic (NR) and SFDC CCSD energy of \ch{Ni(CO)4}
        for different Cholesky thresholds $10^{-x}$, $x=4,5,6,7$
        and non-CD calculations.
        Frozen core of 13 MOs.
        Wall times of the individual steps (in hours) and their contribution (in percent) to the total wall time;
        only one core is used.}
    \begin{tabular}{cccccccccccc}
\toprule
& & $E_{\rm CCSD}$ & \multicolumn{9}{c}{wall time (hrs)}\\
& $x$ & (a.u.) & \multicolumn{2}{c}{ERIs} & \multicolumn{2}{c}{SCF} & \multicolumn{2}{c}{AO2MO} & \multicolumn{2}{c}{CCSD} & total \\
\midrule
NR  & 4	&$	-1960.530412$&	0.8	&	(1.5    \%)	&	0.02	&	(	0.04	\%)	&	0.01	&	(	0.01	\%)	&	56.3	&	(	98	\%)	&	57.1	\\
    & 5	&$	-1960.530424$&	1.2	&	(1.8    \%)	&	0.03	&	(	0.04	\%)	&	0.01	&	(	0.02	\%)	&	64.4	&	(	98	\%)	&	65.6	\\
    & 6	&$	-1960.530416$&	1.4	&	(2.0    \%)	&	0.03	&	(	0.05	\%)	&	0.01	&	(	0.02	\%)	&	71.9	&	(	98	\%)	&	73.4	\\
    & 7	&$	-1960.530416$&	4.0	&	(4.4    \%)	&	0.09	&	(	0.09	\%)	&	0.02	&	(	0.02	\%)	&	87.1	&	(	96	\%)	&	91.2	\\
no CD&	&		 x       &	1.1	&			    &	0.82	&				    &	$>14$	&				&		&				&		\\
no CD AO&&$	-1960.530416$&	1.1	&	(1.5    \%)&	0.87	&	(	1.14	\%)	&	0.75	&	(		\%)	&	73.7	&	(	96	\%)	&	76.4	\\
\midrule														
SFDC& 4	&$	-1973.293766$&	6.1	&	(9.4    \%)&	0.07	&	(	0.11  \%)	&	0.02	&	(	0.03 \%)	&	58.5	&	(	90	\%)	&	64.7\\
    & 5	&$	-1973.293776$&	8.3	&	(11.4   \%)&	0.08	&	(	0.11  \%)	&	0.02	&	(	0.03 \%)	&	64.4	&	(	88	\%)	&	72.8\\
    & 6	&$	-1973.293768$&	9.9	&	(11.9   \%)&	0.09	&	(	0.11  \%)	&	0.03	&	(	0.03 \%)	&	72.8	&	(	88	\%)	&	82.8\\
    & 7	&$	-1973.293769$&	12.8&	(12.9   \%)&	0.11	&	(	0.11  \%)	&	0.03	&	(	0.03 \%)	&	85.9	&	(	87	\%)	&	98.8\\
no CD&	&		  x      &	15.1&			&	4.83	&			    &	$>83$	&				&		    &				&		\\

\bottomrule
    \end{tabular}
    \label{tab:NiCO4-walltime}
\end{sidewaystable}

\begin{table}[htb!]
    \centering
    \caption{Cost of
            the non-relativistic (NR) and CD-SFDC-CCSD \ch{Ni(CO)4} energy calculation
            with unc-ANO-RCC basis set
            for Cholesky thresholds $\cdthr=10^{-x}$, $x=4,5,6$.
            Frozen core of 13 MOs.
            Walltime (total and contributions from the CD and CCSD steps)
            and memory requirements (peak core)
            when 8 OpenMP threads used;
            cf.~Tables~\ref{tab:NiCO4-walltime} and \ref{tab:NiCO4-CD-memory} for single-thread calculations.
            }
    \begin{tabular}{cccccccc}
        \toprule
    & $x$  & \multicolumn{5}{c}{wall time (hrs)} & memory\\
    &   &  \multicolumn{2}{c}{ERIs} & \multicolumn{2}{c}{CCSD} & total & (GB) \\
\midrule
NR  & 4	&	0.3	&	(	3	\%)	&	10.7	&	(	97	\%)	&	11.1		&	157	\\
    & 5	&	0.5	&	(	4	\%)	&	10.2	&	(	95	\%)	&	10.7		&	166	\\
    & 6	&	0.6	&	(	5	\%)	&	13.4	&	(	95	\%)	&	14.1		&	176	\\
\midrule
SFDC& 4	&	1.3	&	(	11	\%)	&	10.8	&	(	89	\%)	&	12.2		&	157	\\
    & 5	&	1.9	&	(	13	\%)	&	13.3	&	(	87	\%)	&	15.3		&	166	\\
    & 6	&	2.3	&	(	15	\%)	&	12.6	&	(	84	\%)	&	15.0		&	176	\\
        \bottomrule
    \end{tabular}
    \label{tab:NiCO4-walltime-8cores}
\end{table}

\begin{table}[htb!]
    \centering
    \caption{Memory requirements (in GB) for
        non-relativistic (NR) and CD-SFDC-CCSD energy calculations
        of the \ch{Ni(CO)4} molecule
        (when one core used),
        cf.~energies and wall times in Table~\ref{tab:NiCO4-walltime}.
        $x$ relates to the Cholesky threshold $10^{-x}$,
        $\ncv$ is the number of CVs,
        in columns CV AO and CV MO is the memory required to store CVs in AO (SCF step)
        and MO (CCSD step) basis set, respectively,
        and in CCSD total is the peak core memory usage during the CD CCSD calculation.
        }
    \begin{tabular}{cccccccccc}
    \toprule
$x$ & $\ncv$ & $f$ & \multicolumn{2}{c}{CV AO} & CV MO & CCSD \\
    & & & NR & SFDC & & \\
    \midrule
4	&	4555 & 113&	4.7	&	9.5	    &	7.6	 &	76.1	\\
5	&	5653 & 91	&	5.9	&	11.8	&	9.4	 &	85.3	\\
6	&	6892 & 75	&	7.2	&	14.3	&	11.5 &	95.6	\\
7	&	8155 & 63	&	8.5	&	17.0	&	13.6 &	106.0	\\
     \bottomrule
    \end{tabular}
    \label{tab:NiCO4-CD-memory}
\end{table}

\begin{table}[htb!]
    \centering
    \caption{The energy of the \ch{Pd(CO)4} molecule
            and the cost of its calculation.
            Non-relativistic (NR) and CD-SFDC-CCSD calculation with unc-ANO-RCC basis set
            for Cholesky thresholds $\cdthr=10^{-x}$, $x=4,5,6$.
            Frozen core of 17 MOs.
            Wall time (total and contributions from the CD and CCSD steps)
            and memory requirements (peak core)
            when 8 OpenMP threads used.
            }
    \begin{tabular}{ccccccccc}
    \toprule
& $x$ & $E_{\rm CCSD}$ & \multicolumn{5}{c}{wall time (hrs)} & memory\\
& & (a.u.) & total & \multicolumn{2}{c}{ERIs} & \multicolumn{2}{c}{CCSD} & (GB) \\
\midrule
NR  &	4	&	-5391.6626	&	0.3	&	(	3	\%	)	&	9.3	&	(	96	\%	)	&	9.6	    &	175	\\
    &	5	&	-5391.6627	&	0.4	&	(	4	\%	)	&	9.7	&	(	95	\%	)	&	10.2	&	185	\\
    &	6	&	-5391.6627	&	0.8	&	(	6	\%	)	&	12.2&	(	93	\%	)	&	13.1	&	195	\\
\midrule
SFDC&	4	&	-5498.0597	&	1.4	&	(	11	\%	)	&	10.5	&	(	88	\%	)	&	11.9	&	175	\\
    &	5	&	-5498.0598	&	2.0	&	(	13	\%	)	&	13.0	&	(	86	\%	)	&	15.1	&	185	\\
    &	6	&	-5498.0598	&	2.5	&	(	17	\%	)	&	11.5	&	(	82	\%	)	&	14.1	&	195	\\
    \bottomrule
    \end{tabular}
    \label{tab:PdCO4}
\end{table}

\begin{table}[htb!]
    \centering
    \caption{The energy of the \ch{Pt(CO)4} molecule
            and the cost of its calculation.
            Non-relativistic (NR) and CD-SFDC-CCSD calculation with unc-ANO-RCC basis set
            for Cholesky thresholds $\cdthr=10^{-x}$, $x=4,5,6$.
            Frozen core of 26 MOs.
            Wall time (total and contributions from the CD and CCSD steps)
            and memory requirements (peak core)
            when 8 OpenMP threads used.
            }
    \begin{tabular}{ccccccccc}
    \toprule
& $x$ & $E_{\rm CCSD}$ & \multicolumn{5}{c}{wall time (hrs)} & memory \\
& & (a.u.) & total & \multicolumn{2}{c}{ERIs} & \multicolumn{2}{c}{CCSD} & (GB) \\
    \midrule
NR  &	4	&	-17785.2227	&	0.3	&	(	3	\%	)	&	9.7	    &	(	96	\%	)	&	10.1	&	204	\\
    &	5	&	-17785.2224	&	0.5	&	(	4	\%	)	&	10.5	&	(	95	\%	)	&	11.0	&	214	\\
    &	6	&	-17785.2225	&	0.6	&	(	6	\%	)	&	10.3	&	(	94	\%	)	&	11.1	&	225	\\
\midrule
SFDC&	4	&	-18875.0728	&	1.3	&	(	11	\%	)	&	10.1	&	(	88	\%	)	&	11.5	&	204	\\
    &	5	&	-18875.0726	&	1.8	&	(	14	\%	)	&	10.5	&	(	85	\%	)	&	12.4	&	214	\\
    &	6	&	-18875.0727	&	2.5	&	(	20	\%	)	&	9.7	    &	(	78	\%	)	&	12.3	&	225	\\
    \bottomrule
    \end{tabular}
    \label{tab:PtCO4}
\end{table}

\begin{table}[htb!]
    \centering
    \caption{Optimized tetrahedral structure of \ch{Ni(CO)4}
            for Cholesky thresholds $\cdthr=10^{-x}$, $x=4,5,6$.
            Non-relativistic (NR) and CD-SFDC-CCSD calculation with unc-ANO-RCC basis set.
            The asterisks denote that the structure could not be converged to higher accuracy due to oscillations.
            The bond lengths are in angstroms.
            Comparison with the experimental values \cite{Hedberg1979}
            and other theoretical works \cite{Frenking2021,McKinlay2015}.}
    \begin{tabular}{ccll}
    \toprule
          & $x$ & r(NiC) & r(CO) \\
    \midrule
      NR  & 4 & 1.8369  & 1.1303 \\
          & 5 & 1.8368  & 1.1302 \\
          & 6 & 1.8368  & 1.1302 \\
      SF  & 4 & 1.8192* & 1.1304*\\
          & 5 & 1.8193  & 1.1304 \\
          & 6 & 1.8193  & 1.1304 \\
      M06-D3/def2-TZVPP~\cite{Frenking2021} &  & 1.848 & 1.132 \\
      CCSD/pVTZ (non-rel.)~\cite{McKinlay2015} &  & 1.831 & 1.147 \\
      Experiment~\cite{Hedberg1979} &  & 1.838(2) & 1.141(2) \\
     \bottomrule
    \end{tabular}
    \label{tab:NiCO4-gopt}
\end{table}

\end{document}